\def \1{\textit{(i)}}
\def \2{\textit{(ii)}}
\def \3{\textit{(iii)}}
\def \4{\textit{(iv)}}
\def \5{\textit{(v)}}
\def \6{\textit{(vi)}}
\def \7{\textit{(vii)}}

\newcommand{\ie}{\textit{i.e., }}
\newcommand{\eg}{\textit{e.g., }}

\newcommand{\cf}{\textit{cf. }}

\newcommand{\sol}{\textit{ThreatFinderAI}}

\documentclass[conference]{IEEEtran}
\usepackage{soul}
\usepackage{tikz}
\usepackage{float}
\usepackage{booktabs}
\usepackage{dblfloatfix}
\usepackage{balance}

\usetikzlibrary{calc}

\begin{document}
\title{Asset-centric Threat Modeling for AI-based Systems}

\author{\IEEEauthorblockN{{Jan von der Assen\IEEEauthorrefmark{1}, Jamo Sharif\IEEEauthorrefmark{1}, Chao Feng\IEEEauthorrefmark{1}, Christian Killer\IEEEauthorrefmark{1}, G\'er\^ome Bovet}\IEEEauthorrefmark{2}, Burkhard Stiller\IEEEauthorrefmark{1}}
\IEEEauthorblockA{{\IEEEauthorrefmark{1}Communication Systems Group CSG, Department of Informatics IfI, University of Zürich UZH}\\
{Binzmühlestrasse 14, CH---8050 Zürich, Switzerland} \\
{E-mail: [vonderassen, cfeng, killer, stiller]@ifi.uzh.ch, jamo.sharif@uzh.ch}}
\IEEEauthorblockA{\IEEEauthorrefmark{2}Cyber-Defence Campus, armasuisse Science \& Technology, CH--3602 Thun, Switzerland gerome.bovet@armasuisse.ch}
}

\DeclareRobustCommand*{\IEEEauthorrefmark}[1]{%
  \raisebox{0pt}[0pt][0pt]{\textsuperscript{\footnotesize #1}}%
}

\maketitle

\begin{abstract}
Threat modeling is a popular method to securely develop systems by achieving awareness of potential areas of future damage caused by adversaries. 
 However, threat modeling for systems relying on Artificial Intelligence is still not well explored. While conventional threat modeling methods and tools did not address AI-related threats, research on this amalgamation still lacks solutions capable of guiding and automating the process, as well as providing evidence that the methods hold up in practice. Consequently, this paper presents \sol{}, an approach and tool providing guidance and automation to model AI-related assets, threats, countermeasures, and quantify residual risks. To evaluate the practicality of the approach, participants were tasked to recreate a threat model developed by cybersecurity experts of an AI-based healthcare platform. Secondly, the approach was used to identify and discuss strategic risks in an LLM-based application through a case study. Overall, the solution's usability was well-perceived and effectively supports threat identification and risk discussion.
\end{abstract}

\IEEEpeerreviewmaketitle

\section{Introduction}
Artificial Intelligence (AI) is considered a disruptive technology that is being integrated into different domains, ranging from healthcare to embedded implementations~\cite{gartner-ai}, which now serve as a key contributor to other technologies such as 6G~\cite{nokia-6g}. In addition to the wide range of domains that show interest in AI technologies, an outstanding observation is the speed at which AI technology is adopted. For example, \textit{ChatGPT} has attracted 100 million monthly active users within weeks~\cite{retuers-chatgpt}. 

The fact that AI technologies are now readily available to individuals, corporations, and national actors has also given rise to concern. For example, \cite{cyd-llm} have analyzed the implications of Large Language Models (LLMs) in the context of the Swiss Cybersecurity landscape, summarizing threats such as spear phishing, vulnerable code injections, and remote code execution~\cite{cyd-llm}. 
Furthermore, researchers have demonstrated that these attacks can be executed in a realistic setting~\cite{engadget-chatgpt}. Aside from LLMs, extensive research has demonstrated potential attacks in related AI technologies, including Machine Learning~\cite{threats-ml}, Federated Learning~\cite{threats-fl}, and Computer Vision~\cite{threats-computervision}.

It appears that the increasing adoption of these technologies and the concerns surrounding them are rightly part of current discussions. However, it is vital to consider that the adoption is ongoing -- organizations are actively integrating these technologies into their products and services. This raises the question of how organizations should approach these security concerns, especially given the scarcity of cybersecurity talent and the speed at which AI services are integrated. 

One approach that has demonstrated value in the conventional application security field is threat modeling, which is used for secure software development, risk assessment, or to foster security awareness~\cite{samm}. 
While threat modeling can help identify and mitigate issues at design time~\cite{coretm}, creating suitable threat models is still challenging for software engineers and data scientists. Multiple reasons can challenge the creation of threat models for AI systems. In research, wide attention is given to investigating threats and vulnerabilities from a research perspective without proposing practical cybersecurity approaches. Furthermore, existing threat modeling methodologies and tools are conceptualized for conventional software systems and, hence, do not directly support AI threat identification. Recent research addressed how to apply threat modeling for AI~\cite{mauri2021stride, wilhjelm2020threat}. However, this limited body of research has not shown how to support or automate the design process. 
Moreover, these approaches were not deployed in scenarios involving real users and design problems.

To fill this gap, the key contributions of this paper is an asset-centric threat modeling and risk assessment approach and a guiding tool. The methodology comprises seven steps that are aligned with the design procedures of AI-based systems. Existing literature is transformed into a queryable knowledge graph to guide and automate threat identification. A stencil library is provided to represent AI-related assets, thereby embedding the semantics of the knowledge graph into diagrams. This supports automated asset and threat identification when modeling AI-based system architectures. Moreover, this work integrates business impact analysis to identify impacts and Monte Carlo simulations using expert-based estimates to quantify residual risk exposure. Finally, experiments demonstrate that the tool can support users in reproducing a threat model created by cybersecurity experts. For this, different types of users were tasked to re-create a threat model based on a given system architecture, followed by a qualitative investigation of the tool's perceived usability. Finally, the cybersecurity experts assessed the user's models. In a second evaluation, the approach's usefulness in a risk analysis scenario is assessed through a case study of an LLM-based architecture for a law firm. Overall, the tool can guide and automate threat modeling for AI and enable subsequent risk analysis. 

\setlength{\tabcolsep}{3pt}
\begin{table}[!b]
\centering
\caption{Literary Work Applying Threat Modeling to AI Systems}
\begin{tabular}{@{}llll@{}}
\toprule
\textbf{\textit{Work}} & \textbf{\textit{Contribution}} & \textbf{\textit{Evaluation}} & \textbf{\textit{Domain}} \\
\midrule
\cite{wilhjelm2020threat} 2020 & Method, Survey & Illustration &  Requirements Engineering \\
\cite{mauri2021estimating} 2021 & Degradation  Method & Demonstration &  Adversarial ML \\
\cite{bitton2022adversarial} 2022 & Threat Model & Demonstration &  Cellular Networks \\
\cite{mauri2022modeling} 2022 & Methodology & Illustration &  AI Threat Modeling \\
\textit{This} 2024 & Methodology, Tool, & Case Study, & AI Threat Modeling \\
 & Risk Quantification & Field Test &  \\

\bottomrule
\end{tabular}
\label{tab:rw}
\end{table}
\setlength{\tabcolsep}{6pt}

This paper is organized as follows. Section~\ref{sec:rw} presents an overview of related literature. While Section~\ref{sec:impl} details the design, evaluations are described in Section~\ref{sec:eval}. Conclusions and directions for future work are outlined in Section~\ref{sec:summary}.

\section{Background and Related Work}\label{sec:rw}

Literature related to this work can be grouped into three segments: \1 research identifying attacks on AI systems, \2 established threat modeling tools and methods, and \3 a small body of literature looking into the combination of the former two. Due to the lack of research on AI threat modeling, painting a realistic picture of the problem domain requires a summary of research in all three areas.

A recent survey organizes cyber attacks on AI systems according to the Machine Learning (ML) pipeline. Data poisoning attacks influence the resulting model by injecting adversarial samples during data collection. These may be falsified at the source or during storage~\cite{mcmahan2017communication}. The attack's goal may vary, and multiple poisoning strategies (\ie random or targeted) exist~\cite{munoz2017towards}. Spanning the feature selection and model training stages, several strategies involve replacing the model with a poisoned one~\cite{sangwan2023cybersecurity}. Attacks achieving model inversion, inference, and failure are described during the inference stage. Model inversion aims to recover information on the training samples~\cite{sangwan2023cybersecurity}, while extraction attacks attempt to obtain or reconstruct the model based on limited access~\cite{sangwan2023cybersecurity}. 

Looking into threat modeling tools and methods, none of the popular tools such as the \textit{Microsoft Threat Modeling Tool}, \textit{CAIRIS}, \textit{Threatspec}, \textit{SDElements}, or \textit{Tutamen} focus on threat modeling for AI systems~\cite{cairis-docs, threatspec, sdelements_datasheet, tutamantic}. Although some present the ability to create custom threat libraries, no taxonomies or AI-related methodologies exist. 

Although industry efforts do not focus on threat modeling as a discipline, several ongoing efforts in describing AI attacks can be found. \textit{MITRE ATLAS} includes attack tactics that are specific to AI~\cite{mitre_atlas}. Another knowledge base that provides a guideline for mitigating AI threats was proposed by Microsoft~\cite{microsoft2023threatmodel}. Similarly, \textit{OWASP} has presented guides to ensure the security of systems relying on AI~\cite{owasp_ai_security}. A comprehensive report of AI-related threats is presented by the European Union Agency for Network and Information System~\cite{enisa_ml} (ENISA). 

The third and most closely related literature group reports evidence of integrating the AI paradigm within threat modeling. The limited number of publications~\cite{mauri2022modeling} (see \tablename~\ref{tab:rw}) connect potential risks to the elements generated throughout various phases of the life cycle of ML models, ranging from the initial requirements analysis to maintenance. 

\cite{wilhjelm2020threat} applies conventional threat modeling consisting of data flow diagramming and STRIDE-based threat identification. While the methodology reports the successful mapping of a threat taxonomy to an illustrative model, experts carry out the mapping process manually. Furthermore, weaknesses such as limited results from a singular synthetic case study and no investigation on usability are acknowledged~\cite{wilhjelm2020threat}.
In \cite{mauri2021estimating}, a gold standard dataset is used to evaluate the degradation of a model during the productive stage. A metric is proposed that quantifies the degradation loss, which could quantify the impact of a threat. However, focusing on existing models might indicate that the method is not applicable during the design stage, making it impractical for architectural threat modeling.
A domain-specific threat model is created in~\cite{bitton2022adversarial}, focusing on Open Radio Access Network (O-RAN) architectures. Thus, no generic approach is evaluated. The paper by~\cite{mauri2022modeling} is the most closely related contribution to threat modeling of AI-based systems. It advocates integrating threat modeling methodologies in AI security analysis and introduces the STRIDE-AI methodology. 
However, it involves a manual mapping process, lacking automation and hindering scalability and adaptability to system changes. The methodology's evaluation is based on a single use case without involving users, providing insights but not covering all challenges in diverse ML applications.

In summary, while one might argue that attacks on AI are not radically different from conventional cyber attacks, it is unclear how straightforward the creation of a threat model for AI is. More specifically, the guiding factors and the degree of automation, especially when creating a threat model by real users during the design stage of a system, is unclear, posing an opportunity for the development of a guiding tool oriented towards the design process of AI system architectures.



\begin{figure*}[t]
    \centering
    \includegraphics[trim={0 .2cm 0 .1cm},clip,width=\linewidth]{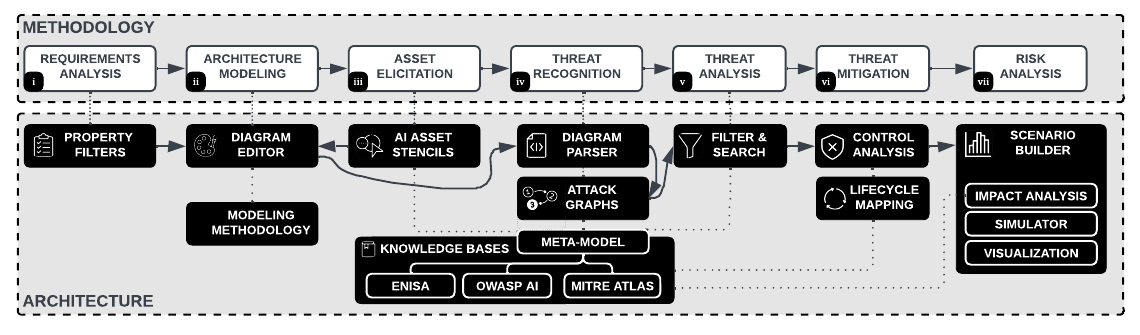}
    \caption{Architecture of the \sol{} Approach}
    \label{fig:arch}
\end{figure*}

\section{The \sol{} Approach}\label{sec:impl}
To design and implement a threat modeling approach for AI-based systems, the architectural semantics of these systems must be mapped to the threat modeling process. The architecture of the \sol{} approach is visualized in \figurename~\ref{fig:arch}. At the top, a high-level overview of the threat modeling procedure is outlined. These steps are aligned with existing research on threat modeling. 
In addition, concepts found in the enterprise risk management (ERM) domain~\cite{hunziker} were adopted, enabling the formulation of strategic risks. 
This process was leveraged to design the individual components that support the overall process and, in sum, provide automated threat modeling for AI-based systems. The architectural components, consisting of eleven key components supporting the approach, are shown below. A prototype was then designed and implemented to investigate the feasibility and effectiveness of the approach. 






\subsection{Methodology and Architecture}
For the objective identification step, literary analysis revealed the necessity to adopt the AI-specific proposal of security principles from~\cite{mauri2022modeling, european2020ai}. There, the traditional CIA principles (\ie confidentiality, integrity, availability) are extended to include authorization and non-repudiation as key concepts. While the definition of important security goals may not be fruitful at this stage, it is crucial to ensure that the business relevance of the system to be developed is well understood~\cite{rcvar}. In the methodology, this step is fulfilled by \1 the scoping of a specific architecture to the scope of a business mission and defining a key security \textit{requirement} and asset. For example, users define a scenario such as ``LLM-based complaint pre-processing'' with ``data'' as an asset and ``confidentiality'' as a key property.

In the second step, the system must be closely analyzed and understood from an architectural perspective. For this, the context for the threat modeling process is essential -- whether for designing a completely new architecture or a threat model is created for an existing system as part of a risk assessment. The methodology behind \sol{} proposes to rely on \2 visual \textit{architectural modeling}. Hence, verifying whether there are existing system diagrams and models is crucial. 
Here, it is essential to draw a holistic picture of the architecture, for which the guiding model of the AI life cycle from~\cite{european2020ai} can be helpful to elicit all activities and the systems involved. Thus, to support the architecture modeling, \sol{} includes a diagram editor and a procedure to ensure that all life stages of the system are incorporated. For example, even when using a pre-trained model, it is important to draw the data collection procedure to capture the whole attack surface, even though a service provider may perform it transparently. A meta-model that abstracts over resources detailing AI attacks is proposed to guide the diagram editor. The key concepts (\ie concepts including processes, environments, data, models) from~\cite{european2020ai} are formalized for architectural modeling. Thus, each item in the diagram is characterized by its category (\eg data, model, procedure, actor, infrastructure). A set of annotated stencils is proposed to ensure that diagrams remain machine-processable, each representing one entity in the meta-model.

\begin{figure*}[t]
    \centering
    \includegraphics[width=.93\linewidth]{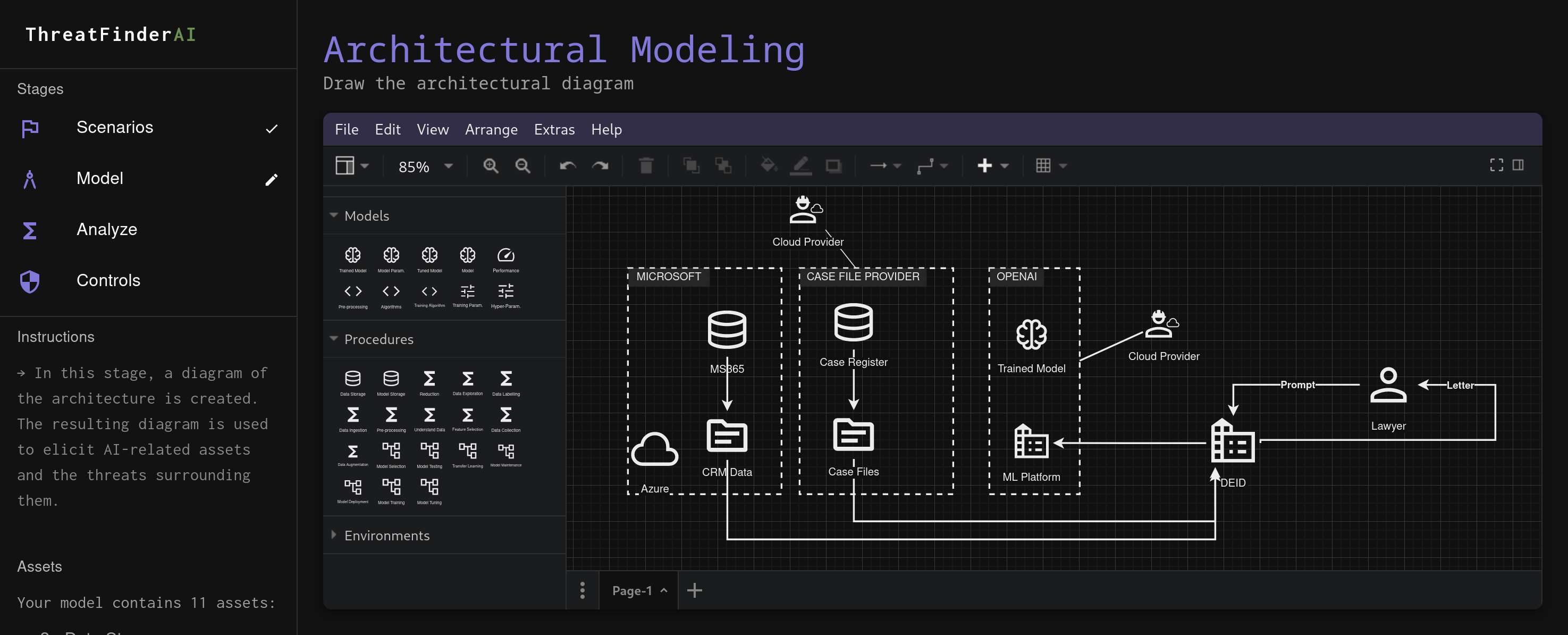}
    \caption{Front end of \sol{}: Architectural Modeling and Asset Annotation Using a Bespoke AI Asset Stencil Library}
    \label{fig:screenshot}
\end{figure*}

In the third step, the architecture model serves as a means to \3 \textit{elicit assets}. Conceptually, these are the functional and data assets subject to the security goals. 
As already mentioned, \sol{} takes a visual approach since it is assumed to be well-established for the development of software architectures. 
Thus, the asset elicitation step can be automated and guided by retrieving the annotations from the architectural diagram and querying them against the previously described meta-model.

The set of assets can then be used as an input for \4 \textit{threat recognition}, yielding related threat events. For example, the presence of training data provided by an untrusted actor could indicate the system's vulnerability to a data poisoning attack. \sol{} achieves threat identification through an attack graph that queries several knowledge bases that are aligned through a graph model. Here, ENISA~\cite{european2020ai}, the OWASP AI Exchange~\cite{owasp_ai_exchange}, and the MITRE ATLAS~\cite{mitre_atlas} catalogues are transformed into a graph-based form. Then, they are related through properties (\eg related asset category, life cycle stage) to the meta-model. Based on this, all knowledge bases can be queried using the previously extracted assets.


Naturally, not all resulting threats are applicable or equally relevant. Thus, the \5 threat analysis phase requires users to navigate the suggestions and add them to a threat model. Here, threats relating to the initially defined \textit{key asset} or \textit{key objective} are highlighted. Following the previous example, the threats indicate that a \textit{data inference} attack can be applied directly to the data or the model. Although users are instructed to review the highlighted key threats (\ie ones targeting the key asset) first, the full list can be filtered based on life cycle, impact, or asset. 
The next step involves \6 
identification of technical, organizational, or strategic \textit{mitigation controls}. 
To automate the step, the knowledge bases can be queried. Here, the meta-model enables control identification through the life cycle stage of the threat. For example, \textit{monitoring a model's usage} is a countermeasure associated with the production stage.

At this stage, a generic threat modeling procedure could be concluded. For example, in a secure software design process, formulating strategic risks may be out of scope, and the key objective is to discover appropriate, practical control mechanisms. However, in a risk assessment context, the output of previous threat models can be reused to \7 analyze strategic risks, thereby generating value through collaboration. 
In modern ERM, all risks should be quantified~\cite{hunziker}. For this, data is needed. Two data sources can be considered -- historical and expert-based. Cybersecurity is still an emerging risk~\cite{hunziker}, and AI security is arguably highly novel within cybersecurity~\cite{mauri2021stride}. Thus, \sol{} proposes to aid experts in quantifying, visualizing, and communicating the uncertainty of AI threats. This is done by identifying business impacts through a set of impact factors compiled from the literature that are mapped to security properties. For example, the previously mentioned data confidentiality breach could lead to breach notification penalties or reputation loss. The following steps in \sol{} facilitate AI threat quantification:
\begin{enumerate}
    \item Build a risk scenario that assesses residual risk by identifying related business impacts. \sol{} provides a set of tangible and intangible business impacts related to a threat's security property. 
    \item For each impact, a distribution of financial impacts can be developed by eliciting $I_{min}$ and $I_{max}$, which represent the worst- and best-case loss estimates. Furthermore, the expert defines a confidence level $I_c$.
    \item The estimated number of occurrences is elicited by defining the best and worst-case number of occurrences (\ie $o_{min}$, $o_{max}$, and $o_c$), including the confidence level.
\end{enumerate}

\begin{figure*}[t]
    \centering
    \includegraphics[trim={0 0 0 5.67cm},clip, width=.95\linewidth]{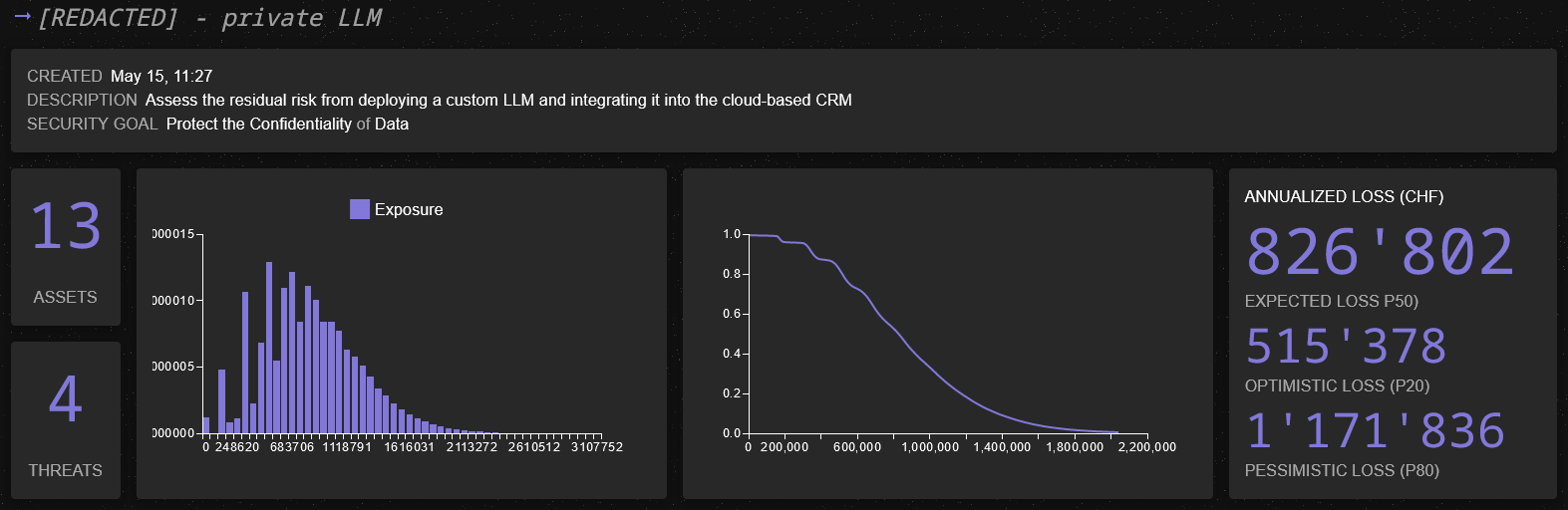}
    \caption{\sol{} Indicating Residual Risk Exposure through Metrics and Visualization of Loss Distribution and Loss Exceedence}
    \label{fig:quant}
    
\end{figure*}
Once these parameters are established, the risk scenarios can be modeled and quantified. In \sol{}, this is achieved by performing Monte Carlo simulations of the two distributions that are built from the experts' estimates. By means of the $I_{min}$, $I_{max}$, and $I_c$, a log-normal distribution of financial losses is built. This distribution is commonly used to model operational losses in risk management~\cite{rcvar}, as it can represent the potential ``long-tail'' across the distribution (\ie more impacts to exceed $I_{max}$ than to fall below $I_{min}$). Since the occurrences of each risk scenario are expressed through discrete values, their distribution is modeled through a Poisson distribution using $o_{min}$, $o_{max}$, and $o_c$.
The exposure distribution is then built by performing the simulations (\ie drawing $n$ values from both distributions). Then, \sol{} presents several metrics and visualizations to assess the residual risk exposure. First, the optimistic ($p=0.2$), pessimistic ($p=0.8$), and expected loss ($p=0.5$) values are displayed. Furthermore, as shown in \figurename{}~\ref{fig:quant}, the probability density function of the losses and the loss exceedance curve can be visualized.

\subsection{Prototype Implementation}
To implement the components for the \sol{} threat modeling approach, the components outlined in \figurename~\ref{fig:arch} were implemented and integrated into a web-based solution. Starting from the front end, the user interacts with a web-based graphical user interface implemented as a Single-page Application (SPA) using React.js~\cite{reactjs}. First, the user creates a new project description and defines the key asset and security goal.

%
By integrating and parametrizing the \textit{diagrams.net} diagram editor~\cite{drawio}, 
the architecture can be modeled without any data leaving the browser. To enable asset elicitation and threat identification, a bespoke stencil library was crafted to simplify and guide the asset modeling stage. The stencil library provides one stencil for each asset identified from the comprehensive report provided by ENISA~\cite{enisa_ml}. 
Adding to the generic threat modeling stencils (\eg data flow arrows, trust boundaries), 72 stencils are formalized into an XML file. This enables the automated annotation of metadata to analyze the resulting diagram. 
Threats are identified by parsing the resulting diagram, extracting assets, and querying any of the knowledge graphs. Each knowledge graph is formatted as a JSON file and aligned by relating threats through the asset-based meta-model that leverages high-level steps and asset categories from~\cite{enisa_ml}. A hierarchical representation enables users to discover threats and controls per asset, evaluate their relevance, and add them to a threat model. 
Finally, residual risks can be quantified by the user as previously described. This is the only time data leaves the browser -- Monte Carlo simulations are performed using the popular Python-based numpy~\cite{numpy} and scipy~\cite{scipy} libraries. After performing $n=100'000$ simulations, the exposure distribution is downsampled and sent back to the front end, where it is visualized (\cf \figurename~\ref{fig:quant}).

The \sol{} prototype only relies on a small number of open-source dependencies (\ie diagrams.net, React, FastAPI). Thus, the full source code is made publicly available~\cite{source-code}, and a running instance is available online~\cite{deployed-tool}.

\section{Evaluations}\label{sec:eval}
\sol{}'s goal is to support, guide, and automate threat identification when modeling threats and risks surrounding AI-based systems. Assessing the effectiveness is challenging since there is a lack of ``perfect'' threat models against which the tool can be tested. Furthermore, threat modeling is still a highly human-centered activity~\cite{threatmodelingmanifesto}. Thus, evaluating \sol{} focuses on whether users could sensibly apply the approach in practical settings, yielding relevant threat models and supporting meaningful risk discussions.

\subsection{Scenario-driven Field Experiment}
In the first evaluation, a scenario-driven field experiment was conducted, which involves both controlled (\ie a clear and repeatable task for participants) and real-world elements (\ie a real AI system architecture and a baseline threat model). First, a threat model is created by security and data science experts collaboratively developing a secure architecture for an AI system in healthcare. Although it cannot be proven that the expertise of the experts leads to an exhaustive and correct threat model, it provides a baseline to compare the threat model created by participants using \sol{}.

The practical context of the model that was created in both steps was the ongoing development of a platform to collect, store, share, and train models from clinical data. The healthcare platform is tailored for clinical data analysis by engineers, clinicians, and researchers. It aims to provide a robust data-gathering system with controlled data synthesis to facilitate experimentation and modeling in the healthcare domain. The platform prioritizes data privacy by incorporating advanced anonymization techniques, attribute-based privacy measures, and reliable tracking systems. The main functionalities of the platform are organized into three primary modules (\ie \textit{Model Training}, \textit{Model Auditor}, and \textit{Data Synthesizer}), seven supporting modules (\ie \textit{Data Anonymization Toolkit}, \textit{Data Uploader}, \textit{Cross-Border Database}, \textit{Dataset Explorer}, \textit{Dataset Builder}, \textit{Dataset Evaluator}, and \textit{Federated Learning}), and three crosscutting modules (\ie \textit{Security Control}, \textit{User Interface}, and \textit{Orchestrator}). 
Data confidentiality and privacy are of utmost importance to business representatives. Not only may data breaches lead to regulatory fines, but the overall trust in the data-sharing platform is a crucial property to stimulate successful data collection.

\subsubsection{Execution}
In the first stage, four experts collaborated to create a threat model of the platform architecture: two with conventional cybersecurity expertise, one with specific AI security knowledge, and the fourth one with a data science background. To do so, they leveraged \textit{diagrams.net}~\cite{drawio} to draw the system architecture, its trust boundaries, potential threat actors, and relevant threats. In total, ten areas of concern were identified and closely investigated. Here, it is important to state that all four experts had to rely on external threat information, which was manually surveyed and compiled using industry reports and academic literature.
Potential system threats were identified and linked to specific threat actors, including malicious platform users, external actors, infrastructure administrators, and automated external entities (\eg bots, malware). These threats span a broad spectrum of security concerns. The experts have pinpointed 44 threats throughout the system. 
In contrast, \sol{}'s attack graph contains 96 threats -- a superset of the threats identified by the experts. 

In the second step, seven participants engaged in a threat modeling workshop using \sol{}, receiving a video-based tutorial on using the tool and information about the previously described platform architecture. 
While the knowledge transfer of the system architecture presents additional complexity, it may not be unrealistic for a threat modeling scenario (\eg when onboarding a cybersecurity expert). 
Once participants applied the tool to the scenario, the threat models were collected, and the participants were guided through a questionnaire to understand the perception of  \sol{}.

First, the background and expertise of the participants were elicited. As visible from \tablename~\ref{tab:participants}, participants from different backgrounds were selected. None of the participants indicated cybersecurity knowledge, and only two had a Computer Science degree in Data Science. These participants were the only ones who stated practical knowledge of working with AI. The remaining participants considered themselves theoretically knowledgeable.
Next, it was investigated whether the assumption that IT professionals are familiar with the \textit{diagrams.net} editor was justified. Based on statements shared on a Likert scale, all technical participants expressed familiarity. None of the participants expressed awareness of threat modeling tools.
\begin{table}[t]
    \centering
    \caption{Participants Background and Usability Rating}
    \label{tab:participants}
    \begin{tabular}{@{}cll|l@{}}
        \toprule
        \textbf{\#} & \textbf{\textit{Educational Background}} & \textbf{\textit{AI Knowledge}} & \textbf{\textit{SUS Score}} \\
        \midrule
        1 & Master of Data Science & Practical experience & 55 \\
        2 & Master of Data Science & Practical experience & 70 \\
        3 & B.Sc. Software Systems & Theoretical knowledge & 85 \\
        4 & B.Sc. Software Systems & Theoretical knowledge & 52.5 \\
        5 & B.Sc. Information Systems & Little to no understanding & 52.5 \\
        6 & M.Sc. Pharmacy & Little to no understanding & 75 \\
        7 & Master of Law & Little to no understanding & 45 \\
        \bottomrule
    \end{tabular}
\end{table}
Additional questions investigated the perceived ability to use the tool. All participants felt successful in navigating the tool. Participant Six (\ie the pharmacist) faced challenges during asset identification, acknowledging a limited understanding of AI system architecture. When rating the clarity of the task instructions, six out of seven considered them sufficiently clear. All participants expressed confidence about their understanding of the architecture and scenario. 
Concluding the questionnaire, the perceived usability was assessed using the system usability scale (SUS), providing a simple, standardized scoring using ten questions~\cite{gitlab-handbook}. The resulting scores are shown in \tablename~\ref{tab:participants}. Considering all participants, the average score evaluates to \textit{acceptable usability}. Out of the participants with a computer science background, Participant One's score stands out negatively, while the individual responses conflict. For example, the participant indicated that he/she might \textit{require assistance from a technical person} while expressing that the tool is \textit{easy to learn}. Based on additional open feedback collected, Participant One disliked the diagram editor leveraged in \sol{} while positively acknowledging the platform's features.

\subsubsection{Analysis}
Since the threat modeling process is guided and automated, artifacts, such as the annotated architecture diagrams and the threat models, were recorded and analyzed by the initial threat analysis experts. In the objective selection, it was observed that the participants and the expert group agreed on the security properties and assets. 
Based on the expert assessment, Participants One, Two, Three, and Six successfully identified all relevant threats, underscoring the advantage of AI knowledge in threat modeling for AI-related systems. While this does not come as a surprise, given that they are the most relevant target group, it is surprising that even a layperson achieved this. 
Therefore, the participants with a data science background discovered all relevant threats (and more granular variants) from the expert-based model. While the application is effective in helping users discover relevant threats, users are still required to discard less relevant threats. 


In summary, participants effectively identified all potential threats using the architectural model, addressing previously identified gaps in the literature. However, efficiency could be improved by further filtering threats, potentially through subcategories of security objectives and requirements. The experiment shows that practitioners with a technical background could use this tool as guidance for an initial threat identification step. The resulting threat model would require further analysis either by leveraging additional tools, involving security experts, or conducting cybersecurity training.

\subsection{Case Study: Risk Identification and Quantification}
In the second experiment, \sol{} is applied in a real-world context through a case study within a legal advisory company. One of their services entails the analysis of a case through a legal advisor, which frequently leads to the creation of an objection letter. The creation of such letters is highly repetitive: after manual inspection, the advisor tasks an assistant to parameterize a template using customer and case information. Since the company already stores its customer data in a cloud-based Customer Relationship Management system (CRM), it is intrigued whether integrating an LLM-based architecture could automate this process. The following questions were guiding the case study. 
\begin{enumerate}
    \item[\textbf{\textit{Q1)}}] Which threats arise for such an integration?
    \item[\textbf{\textit{Q2)}}] Which countermeasures can be considered?
    \item[\textbf{\textit{Q3)}}] Does the integration pose a strategic risk? 
\end{enumerate}
\subsubsection{Execution}
First, a new scenario is created within the platform. Then, together with the domain expert, \textit{confidentiality} is defined as a key goal, and customer \textit{data} as a key asset. Since the advisor would perform manual checks of the generated letters, the integrity of the resulting letters is not as critical. Furthermore, although the automation would potentially improve efficiency, the system's availability is not critical since the letters can still be created manually.

Once these high-level requirements were established, a potential integration architecture was developed. 
Instead of the domain expert, a subject-matter expert on cloud-based systems (hereinafter referred to as ``architect'') developed an integration architecture. Since no previous architecture diagram existed, a new one was drawn. The architect leveraged the stencils provided by \sol{}. The resulting architecture involved two cloud platforms (one hosting the existing CRM and the other hosting a service-based LLM). In addition, a web API of a third party provides access to case files. Due to data privacy concerns, the company did not allow the publishing of the architecture; however, the developed architecture represents a more detailed variation of the one indicated in \figurename{}~\ref{fig:screenshot}.

Once the architecture was drawn, the architect navigated to the ``analyze'' page, where each asset was automatically extracted from the diagram. Although the architect is unfamiliar with the \textit{OWASP AI Exchange} knowledge base, it was selected due to familiarity with other OWASP initiatives. \sol{} automatically identified ten assets. Then, due to the initial objective definition  (\ie \textit{data confidentiality}), only four assets are displayed: \textit{customer information}, \textit{case files}, \textit{prompt input} and \textit{output}. Since the case files are from a third-party service, they are not directly critical to the company. Since \textit{customer information} is clearly the key asset to be protected, it was further analyzed. \sol{} suggested six threats that relate to the \textit{confidentiality} of the data assets. To filter the threats, development-time threats were excluded through the life cycle filter (\ie ignoring model development stages), since the customer data is not used for model refinement or training. Thus, the following three key threats remained from the OWASP-based knowledge base.
\begin{enumerate}
    \item[{\textbf{T1}}] Leaking sensitive input data, through the prompt.
    \item[{\textbf{T2}}] Sensitive data in the output (\eg copyrighted text or customer information).
    \item[{\textbf{T3}}] Data exfiltration in the CRM where the data is stored.
\end{enumerate}
Although all three are relevant and important to consider to secure the architecture holistically, the third threat \textit{T3} is already known since the customer data is already stored in the cloud. Since the objective of the case study is to identify the strategic risks of adopting the LLM, the first two threats are prioritized.
The control identification is carried out after adding the two threats to the threat model. Here, the knowledge base is queried using two properties: the life cycle stage (\ie model usage) and the asset type (\ie data assets), yielding the control measures highlighted in \tablename{}~\ref{tab:controls}.

\setlength{\tabcolsep}{6pt}
\begin{table}[b]
    \centering
    \caption{Controls Identified for the Two Key Threats}
    \label{tab:controls}
    \begin{tabular}{@{}lll@{}}
        \toprule
        \textbf{\textit{Threat}} & \textbf{\textit{Countermeasures}} \\
        \midrule
         & Data Minimization and Anonymization or Deidentification  \\
        \textit{T1} & Control Access to Customer Info  \\
         & Encrypting the Communication Channel  \\
         \midrule
         & Detect and Filter Sensitive Fields  \\
        \textit{T2} & Output Monitoring and Retention  \\
         & Manual Review of Output  \\
         
        \bottomrule
    \end{tabular}
\end{table}

Apparently, data minimization and anonymization techniques, as well as verifying the output of the LLM, are critical controls. Thus, the architecture should be revised to include \1 automatic anonymization during prompt creation by an automated middleware service, \2 logging input and output, and \3 reviewing the output before using it as a letter.

Based on a discussion of insights from the company, all control measures were considered feasible. Nevertheless, not all controls would be perfectly effective (\ie completely mitigate the threats). Thus, two residual risks remain after applying the impact analysis function of \sol{}. First, an anonymization failure could lead to the leakage of personal data, thus breaching customer trust and privacy regulations. The platform proposed two business impacts: data protection fines and potential customer losses. For the second risk, the failure to recognize the presence of copyrighted material in the resulting letter could lead to a legal dispute since the case files are provided by a commercial service.
\setlength{\tabcolsep}{8pt}
\begin{table}[t]
    \centering
    \caption{Impact Estimates over Ten Years}
    \label{tab:mc}
    \begin{tabular}{@{}l|rrrl@{}}
        \toprule
        \textbf{\textit{Scenario}} & \textbf{\textit{Occurrence}} &  \textbf{\textit{Losses (CHF)}} & \textit{\textbf{Confidence}} \\
        \midrule
        Data Protection Fine  & $[0, 2]$ & $[1E4, 1E5]$ & 0.9 \\
        Customer Losses  & $[0, 5]$ & $[1E5, 5E5]$ & 0.9 \\
        Legal Dispute  & $[0, 1]$ & $[2E5, 3E5]$ & 0.95 \\
         
        \bottomrule
    \end{tabular}
    Occurrences Are Modeled Over a 10-Year Horizon, Losses are Per Event
\end{table}

The impacts were then quantified using the risk simulator in \sol{}. There, the tool gathered the estimated lower and upper bound for each risk, including the probability of occurrence, loss, and confidence level. Two aspects were considered: the applicability of Swiss privacy law with potential fines and the adoption of an insurance product. Then, the losses and occurrences displayed in \tablename{}~\ref{tab:mc} are estimated.
After \sol{} performed 100'000 Monte Carlo simulations, the annualized loss expectation was visualized, as summarized in \tablename{}~\ref{tab:exposures}. Based on the range of potential losses, there was high uncertainty regarding the impacts. Furthermore, when interpreting the overall risk exposure (\ie aggregating all scenarios), it appears that the introduction of the LLM-based architecture could pose a strategic risk if the company is not willing to formulate a corresponding risk appetite. If the system were adopted, \sol{} indicates the exposure per threat. Hence, non-technical measures can be considered, too. For example, the law firm could offer their customers a transparent opt-in-based service, where they decide whether to pay for a semi-automated or a pricier, fully expert-driven service. This could reduce the impact of customer losses since customers would be aware of the technology's usage.

\setlength{\tabcolsep}{10pt}
\begin{table}[H]
    \centering
    \caption{Simulated Losses (in CHF) Over Ten Years}
    \label{tab:exposures}
    \begin{tabular}{@{}l|rrrl@{}}
        \toprule
        \textbf{\textit{Scenario}} & \textbf{\textit{Optimistic}} &  \textbf{\textit{Expected}} & \textit{\textbf{Pessimistic}} \\
        \midrule
        Data Protection Fine  & 69'162 & 139'647 & 233'629 \\
        Customer Losses  & 1'177'150 & 1'822'151 & 2'667'088 \\
        Legal Dispute  & 0 & 292'027 & 582'141 \\
         
        \bottomrule
    \end{tabular}
  Based on exceedance scenarios with $P=0.2$, $P=0.5$, and $P=0.8$
\end{table}

\subsubsection{Findings}
Based on the application of \sol{}, the initial questions of the case study can be discussed. Regarding \textit{\textbf{Q1}}, the \sol{} approach successfully enabled the identification of relevant threats by modeling the architecture of a system. Although the threats cannot be proven to be complete, the approach was successfully applied to an architectural design to stipulate threat identification. In the case study, an LLM-based architecture was analyzed. Although the methodology was developed for generic AI architectures (\ie not specifically for LLMs), the proposed meta-model unifying the different knowledge bases is generic enough to support this architecture. Although domain experts have to apply their expertise to build a threat model using the suggested threats, \sol{} automatically suggested several relevant countermeasures, effectively answering \textit{\textbf{Q2}}. For example, as shown in \tablename{}~\ref{tab:controls}, data minimization arises as a constructive control to be implemented. Finally, \sol{} was used to elicit potential business impacts. These impacts were then quantified in an expert-based simulation. Due to the lack of data on AI risks, it is impossible to prove the estimations' correctness. However, it can be argued that \sol{} supports formulating strategic risks. For example, the previously discussed exposures could facilitate the discussion that customer losses pose a strategic risk for the company. Moreover, experts can use the visualizations to express their degree of uncertainty by defining the confidence level and the distance between optimistic and pessimistic losses. Thus, the proposed business impact analysis and quantification present successful answers for \textbf{\textit{Q3}}.

\section{Summary and Future Work}\label{sec:summary}
Due to the necessity of considering AI-based system architectures and their security concerns, this paper proposes \sol{}. The approach aligns AI security with threat modeling through a guiding approach and prototype. The prototype includes a front end to define security objectives and enables architectural diagramming with a bespoke stencil library of AI assets. The diagrams are automatically analyzed against an aggregated knowledge graph spanning industry reports. In addition, \sol{} enables control identification, business impact analysis with a set of compiled impacts, and the ability to perform and visualize risk quantification.

To understand the prototype's practicability, effectiveness, and usability, a small-scale user experiment confronted users with a real-world AI system architecture. The results show the feasibility of the AI threat modeling approach, as well as the effectiveness when non-security experts identify threats. Furthermore, the case study of an LLM-based system demonstrates that the approach can be deployed in a real-world scenario. Here, \sol{} automatically identified key threats and countermeasures and helped communicate the residual risk exposure. Hence, based on these results, it is concluded that \sol{} presents a valuable contribution to the security management of AI-based systems, especially when integrating it directly into the design phase of a system. Here, it could serve as a tool to help security engineers automate the threat modeling stage or help non-security experts adopt threat modeling when building AI systems.
In the future, the usability and effectiveness will be further tested with a broader set of participants and scenarios. Moreover, the approach will be tailored to specific AI trends, such as Federated Learning. 


\section*{Acknowledgments}
This work has been partially supported by \textit{(a)} the Swiss Federal Office for Defense Procurement (armasuisse) with the CyberMind and RESERVE (CYD-C-2020003) projects and \textit{(b)} the University of Zürich UZH.

\balance
\bibliographystyle{IEEEtran}
\bibliography{bib/main.bib}
\noindent \small{\\All links above were last accessed on \today{}}

\end{document}